\documentclass[reprint,nofootinbib]{revtex4-1}

\usepackage[english]{babel}
\usepackage[utf8x]{inputenc}
\usepackage[T1]{fontenc}

\newcommand{\Abl}[2]{\frac{{\rm d} #1}{{\rm d} #2}}

\newcommand{\bsub}{\begin{subequations}}

\newcommand{\esub}{\end{subequations}}

\newcommand{\beal}{\begin{align}}

\newcommand{\enal}{\end{align}}

\newcommand{\bspl}{\begin{split}}

\newcommand{\espl}{\end{split}}

\newcommand{\nul}{$\nu_{\scriptscriptstyle{\mathrm{L}}}$}
\newcommand{\nuz}{$\nu_{\scriptscriptstyle\mathrm{z}}$}

\newcommand{\kk}{\bdv{k}}
\newcommand{\bdv}[1]{{\bf #1}}

\newcommand{\ion}[2]{{\text{{\sc #1}\,{\sc #2}}}}

\usepackage{amsmath}
\usepackage{enumitem}
\usepackage{graphicx}
\usepackage[colorinlistoftodos]{todonotes}
\usepackage[colorlinks=true, allcolors=blue]{hyperref}
\usepackage{cleveref}

\begin{document}

\title{Revisiting Primordial Black Holes Constraints from Ionization History}
\author{Benjamin Horowitz}
\email{bhorowitz@berkeley.edu}
\affiliation{Department of Physics, University of California at Berkeley, Berkeley, CA 94705}

\begin{abstract}
Much attention has been drawn to the recent discoveries by the Advanced Laser Interferometer Gravitational-Wave Observatory (LIGO) of merging intermediate mass black holes. Of particular interest is the possibility that the merger events detected could be evidence of dark matter in the form of primordial black holes (PBHs). It has been argued that the presence of many black holes would effect the thermal and ionization history of the universe via their accretion of matter which would have strong signatures in the Cosmic Microwave Background's (CMB) power spectra evident in the damping of anisotropies and change in low-$l$ polarization power. In general the accretion is quite sensitive to the specific physics involved and the conditions of the early universe. In this work, we take a minimal approach and find constraints on PBHs not including the model dependent effects of nonlinear structure of formation or transition between different accretion models which would work to increase the effect. In addition, we include the relative velocity between dark matter and baryonic matter including the effects of supersonic streaming at high redshift which work to significantly reduce the constraining power. We also examine the constraints on more astrophysically-motivated extended black hole mass functions and discuss how mergers might effect this distribution. We find constraints on PBHs in the range $ \approx 30 M_\odot$, finding that they could not compose more than $10\%$ of the total dark matter content.
\end{abstract}
\maketitle
\section{Introduction}

The formation of primordial black holes (PBHs) is a common feature in many theories of early universe, both with \cite{carr1993primordial,ivanov1994inflation,yokoyama1995formation} and without \cite{carr1974black, bicknell1979formation} inflation. PBHs above $10^{-19} M_\odot$ would be sufficiently massive to not evaporate via Hawking radiation by the present day and they have been a dark matter candidate for some time \cite{negroponte1980polarization}. The recent detection of merging $ \sim 30 M_\odot$ black holes from the Advanced Laser Interferometer Gravitational-Wave Observatory (LIGO) has added significant new interest in this possibility due to the seemingly high merger rates indicative of a large population in this mass range \cite{kashlinsky2016ligo,bird2016did,sasaki2016primordial,clesse2016detecting}.

Over the past decades, various studies have found constraints on the mass of PBHs over a broad range using a variety of astrophysical probes \cite{green2015primordial,carr2016primordial}. Of particular note in the LIGO mass range has been the surveys of the MACHO \cite{alcock2000macho} and EROS teams \cite{tisserand2007limits} which have used searches for micro-lensing events in the Local Group to constrain number of black holes in the galactic halo. More recently, constraints from dynamical friction within ultra faint galaxies have excluded a broad range of potential PBH parameter space\cite{brandt2016constraints}. Going forward, new detection techniques from pulsar timing \cite{schutz2016pulsar}, lensing of fast radio bursts \cite{munoz2016lensing}, and disruption of wide binaries \cite{monroy2014end} may further solidify constraints.  

All probes have some model-dependency and even the more well-tested constraints could be weaker than usually presented. In the case of microlensing studies, it has been argued that the rate of events is sensitive to the specific dark matter profile of the Milky Way\cite{hawkins2015new}. Similarly, \cite{brandt2016constraints} admits that the dwarf galaxy heating constraints could be weaker if there happens to be an otherwise undetected intermediate ($10^4 M_\odot$) black hole stabilizing the system and providing binding energy. This is not to suggest a fundamental concern with any given technique, but to motivate the interest in exploring additional independent probes to be more confident of existing constraints. 

Similarly, it is important to examine the effects of extended mass functions on constraints \cite{green2016microlensing}. As most probes have a characteristic mass scale of sensitivity, it is useful to understand the constraining power both within and outside this range. As many astrophysical motivated mechanisms for PBH formation expect an extended distribution for the black hole mass, it is natural to constrain these types of models as well \cite{carr2016primordial}. 

As noted by \cite{ricotti2008effect}, a large population of sufficiently massive accreting PBHs would create significant changes in the CMB, both in terms of anisotropies and spectral distortions. Using WMAP3 and FIRAS data, strong constraints were found in the mass range of $1 M_\odot < M_{pbh} < 10^8 M_\odot$. Later work has suggested weaker spectral distortion constraints \cite{clesse2016detecting}, and, as noted by \cite{bird2016did}, both these constraints depend significantly on particular assumed gas physics around the PBHs and the formation of large scale structure in the early universe. Of particular note is the effect of relative velocity between dark matter and baryonic matter \cite{tseliakhovich2010relative} which works to suppress creation of accretion zones around black holes at high redshift.

In addition, since the work of \cite{ricotti2008effect}, there has been new understanding of the effect of energy deposition in the ionization fraction and thermal history of the universe coming from interest in dark matter annihilation and decay models \cite{slatyer2013energy,madhavacheril2014current} which hasn't yet been applied to black hole accretion's effects on the Intergalactic Medium. 

In this work we revisit the CMB constraints on primordial black holes taking a conservative approach in the dynamics of the accretion. In particular, we will assume no particular nonlinear large scale structure history and an inefficient accretion model with realistic energy deposition into the IGM. In Section \ref{PBHP}, we will introduce our black hole accretion model, which is a limiting case of that discussed in \cite{ricotti2008effect}, and our mass functions under consideration. In Section \ref{IGM}, we discuss our energy deposition and ionization model with particular focus on its differences to those discussed in \cite{ricotti2008effect,chen2016constraint}. In Section \ref{CMB_An}, we discuss the effect on the CMB of the change in ionization history and the parts of the CMB power spectra driving our constraints. In Section \ref{Discussion}, we discuss our results and their implications on constraints on PBHs.

\section{Primordial Black Hole Physics}
\label{PBHP}
\subsection{Accretion Physics}
\label{Accretion_Physics}
The two main regimes of accretion are spherical accretion (aka Bondi accretion \cite{bondi1944mechanism}) and disk accretion. These two regimes are distinguished by the angular velocity of the incoming gas which, in general, depends on many nonlinear effects around the black hole environment. However, as discussed in \cite{mack2007growth}, in all regimes spherical accretion will be less efficient due to feedback effects of thermal pressure within the gas envelope around the black hole. We will assume all accretion is in the spherical regime to provide a firm lower bound. We will often work in terms of the Eddington Mass Rate and Luminosity, defined as 
\bsub
\beal
\label{edd}
 L_{Ed} \equiv & 1.3 \times 10^{38} (\dot{M}_{pbh}/M_\odot) \textrm{ erg s}^{−1},
\\
\label{eddington_lum}
 M_{Ed} \equiv & L_{Ed}/c^2 =
1.44 \times 10^{17}(M_{pbh}/M_\odot) \textrm{ g s}^{−1},
\end{align}
\esub
As discussed in \cite{ricotti2008effect, mack2007growth} the luminosity outflow, $L$, of a primordial black hole is proportional to the square of the dimensionless mass inflow $\dot{m} = \dot{M}/\dot{M_{Ed}}$.
\bsub
\beal
\label{bondi_lum}
 L =& 0.011 \dot{m}^2 L_{Ed},
\end{align}
\esub
In the limit that primordial black holes dominate the dark matter distribution, we can find the mass infall rate as
\begin{equation}
\dot{m} = (1.8 \times 10^{-3} \lambda) \left(\frac{1+z}{
    1000}\right)^3 \left(\frac{M_{pbh}}{1
  M_\odot}\right)\left(\frac{v_{eff} }{5.74~{\rm km
    ~s}^{-1}}\right)^{-3},
\label{eq:acc1}
\end{equation}
where the prefactor $\lambda$ depends on the gas viscosity and is defined as in \cite{mack2007growth} (section 3.5) and $v_{eff}$ is the effective gas velocity, depending on the relative velocity between dark matter and bayrons $\langle V_{rel}\rangle$ and the sound speed of the gas $c_s$. This relative velocity terms requires some care due to the delay in decoupling between dark matter and baryons which works to create fast baryonic flows, so callled ``supersonic streams,'' in the dark matter \cite{tseliakhovich2010relative,yoo2011supersonic}. The relative velocities $\bdv{V}_{rel}$ of the baryon and dark matter distributions can be found in fourier space by 
\begin{equation}
\bdv{V}_{rel}(\kk,z)=\frac{\hat{\kk}}{ik}(\theta_{\text{cdm}}(\kk,z)-\theta_{\text{bar}}(\kk,z)),
\label{eq:velo}
\end{equation}
where $\hat{\kk}$ is the unit vector and $\theta$ is the velocity divergence. The divergence terms are computed using the Boltzmann code CLASS \cite{lesgourgues2011cosmic} and the average relative velocity is found by
\begin{equation} \label{eq:vbc}
\langle V_{\rm rel}^2(\vec{x}) \rangle = \int \frac{dk}{k} \Delta_\zeta^2(k)\left[ \frac{\theta_b (k) - \theta_c (k) }{k}\right]^2,
\end{equation}

where $\Delta_\zeta^2(k) = 2.42\times10^{-9}$ is the initial curvature perturbation variance per $\ln k$. For a Planck Cosmology at recombination, we find $\langle V_{\rm rel}(\vec{x}) \rangle \approx$ 31 km s$^{−1}$. This velocity term is significantly higher than that found in \cite{ricotti2008effect} which didn't account for differential decoupling time, finding a relative velocity of $\langle V_{\rm rel}(\vec{x}) \rangle \approx$ 3 km s$^{−1}$ at recombination. We label the later case the ``no supersonic streaming" case to show the relative effects on constraints.

Again following \cite{mack2007growth}, defining the ``cosmic Mach number" as ${\cal K}_{pbh} \equiv \langle V_{rel}\rangle /c_s$ we have
\begin{equation}
\langle v_{eff} \rangle \approx
\begin{cases}
c_s  \left[{{16 \over \sqrt{2\pi}}{\cal K}_{pbh}^{3}}\right]^{1 \over 6} & \textrm{for }{\cal K}_{pbh}>1, \cr
c_s(1+{\cal K}_{pbh}^2)^{1 \over 2} & \textrm{for }{\cal K}_{pbh}<1. \cr
\end{cases}
\label{eq:alpha_6}
\end{equation}

For the range of $M_{pbh}$ and $v_{eff}$ under consideration, $\dot{m}<0.1$ meaning the luminosity of a given PBH would be sub-Eddington.

In order to understand the effects of this energy source on the thermal history of the universe, we need to model the output spectra of the accretion zone. The specific case of Bondi accretion was studied in \cite{shapiro1973accretion}, showing that the spectrum is $\nu L_\nu \propto \nu^{0.5}$ with an exponential cutoff at $\nu_\text{cut} \sim 5 \times 10^5 \textrm{ eV}$. Like the results found for lower redshift quasars, this spectrum is ionizing and also deeply penetrating and we expect to see similar outside-in ionization patterns as found in \cite{haiman2001photon} so we can largely ignore the specific optical properties of the accretion halo.

The heat of an accreting black hole will create an outward pressure on the surrounding mass, reducing the matter infall rate temporarily until the matter cools and the pressure decreases. The effect is controlled by the duty cycle parameter, $f_{duty}$ of the black hole. 
However, \cite{milosavljevic2009accretion2} has argued via scaling arguments that the duty cycle should be roughly equal to one for black holes in the mass range $1 M_\odot < M < 100 M_\odot$. This result is also supported by simulations by \cite{mack2007growth}.
With this wide range in mind, we present our constraints in terms of $f_{duty}$ and use a fiducial value $f_{duty}=1.0$ for $M_{pbh}<100 M_\odot$. In general $f_{duty}$ will be sensitive to turbulent gas physics around the black hole and studies (both simulations and observations) \cite{shapley2003rest,milosavljevic2009accretion,park2011accretion,park2012accretion} have found rates ranging from 2\% to 33\% for supermassive black holes, like those powering quasars. We use a conservative value of $f_{duty} = 0.02$ for constraints involving mass functions extending into $M>100 M_\odot$.

\subsection{Other effects}
\label{other_effects}
A natural question to consider is how the growth of structure would effect these results. There have been arguments suggesting that PBHs frequently form binary systems or clump in small compact halos \cite{dokuchaev2004quasars,clesse2016clustering}. While this was considered in \cite{ricotti2008effect}, the results depend significantly on the specifics of the structure of formation and the mechanics of black hole accretion. Formation of halos at high redshift and effect on relative velocity was studied in \cite{stacy2011effect}. The formation of halo structure was found to increase the relative velocity at late times. However, as discussed in \cite{ricotti2008effect}, the formation of dark matter halos will change the dynamics of the black hole accretion zones and change the accretion mechanism to disk-like accretion, which is more luminous and can approach the Eddington limit.

Similarly, mergers of black holes during the cosmic dark ages will result in a skewing of the primordial black hole distribution to higher masses which have proportionally higher luminosity since $l \propto M_{pbh}^2$. The exact rate of black hole mergers during this era depends on the initial conditions of the PBHs which will determine the formation of early binaries \cite{nakamura1997gravitational,clesse2015massive} and how early clusters form \cite{tseliakhovich2010relative}. We leave a more detailed analysis of the merger rate during this time-period to future work.

Other effects to consider include perturbative effects of magnetic fields and cosmic ray production inside the accretion zone. These effects have been shown to be small and primarily would increase accretion rate \cite{meszaros1989radiation,mack2007growth}.

\subsection{Mass Functions}

For our analysis, we will consider two black hole mass functions \cite{carr1975primordial}. To make contact with existing constraints, our ``baseline'' model will use the familiar delta function used in most constraints so far. This function is parametrized by the fraction $f_{pbh}=\Omega_{pbh}/\Omega_{cdm}$ and the location of the peak $M_{pbh}$. However, there is no particular reason to believe the PBH mass function should be so focused, and there are various physical reasons to believe that primordial black holes would follow an extended spectrum \cite{niemeyer1998near}. Following the dynamic and weak lensing constraint of \cite{green2016microlensing}, for our ``extended'' mass function use a form;
\begin{equation}\label{eq:dmdn}
  \frac{dn}{dm}= N \exp{\left( \frac{(\log{M}-log{M_c})^2}{2\sigma_{pbh}^2}\right)},
\end{equation}
where $N$ is a normalization constant, $\sigma_{pbh}$ is a distribution of masses, and $M_c$ is the peak of the distribution. This form is particularly interesting as it is able to capture most of the behavior of PBH distributions arising from axion-curvaton and running-mass inflation \cite{green2016microlensing}. For the purpose of our analysis to keep our parameter space reasonable, we will look only at this model where $f_{pbh} = 1$. While these model have been ruled out by the analysis of \cite{green2016microlensing}, we revisit it to demonstrate the relative constraining power of the CMB analysis.

As discussed in \cite{carr2016primordial}, it should be possible to construct constraints for any given extended mass function from a dirac delta mass function constraint through use of weighting of the appropriate mass bins. However, as mentioned in \cite{green2016microlensing}, this method seems to systematically underestimate constraints due to edge effects of given bins. Even if these differences are small, it might be useful to consider the effect of extended mass functions if one wants to understand the evolution of the mass function and compare constraints originating from different redshifts.

\section{IGM and Ionization History}
\label{IGM}

The IGM's ionization history is parametrized by the Thomson scattering optical depth defined by
\begin{equation}\label{eq:tau}
  \tau(z,z_0) = \int_{t(z)}^{t(z_0)} n_{\rm e} \sigma_{\rm T}\,c\mathrm{d}t' ,
\end{equation}
where $n_{\rm e}$ is the number density of free electrons at time $t'$,
$\sigma_{\rm T}$ is the electron-photon scattering cross-section, and $t(z)$ is the time at redshift $z$. As the number density changes with cosmic expansion, it is often easier to work in terms of the ionization fraction, $x_{\rm e}(z) \equiv n_{\rm e}(z)/n_{\rm H}(z)$, where $n_{\rm H}(z)$
is the hydrogen number density. Depending on the recombination/reionization model, this fraction will range from $10^{-4}$ to a bit above 1 (indicative of ionization of helium in addition to hydrogen). 

\begin{figure}
    \includegraphics[width=0.50\textwidth]{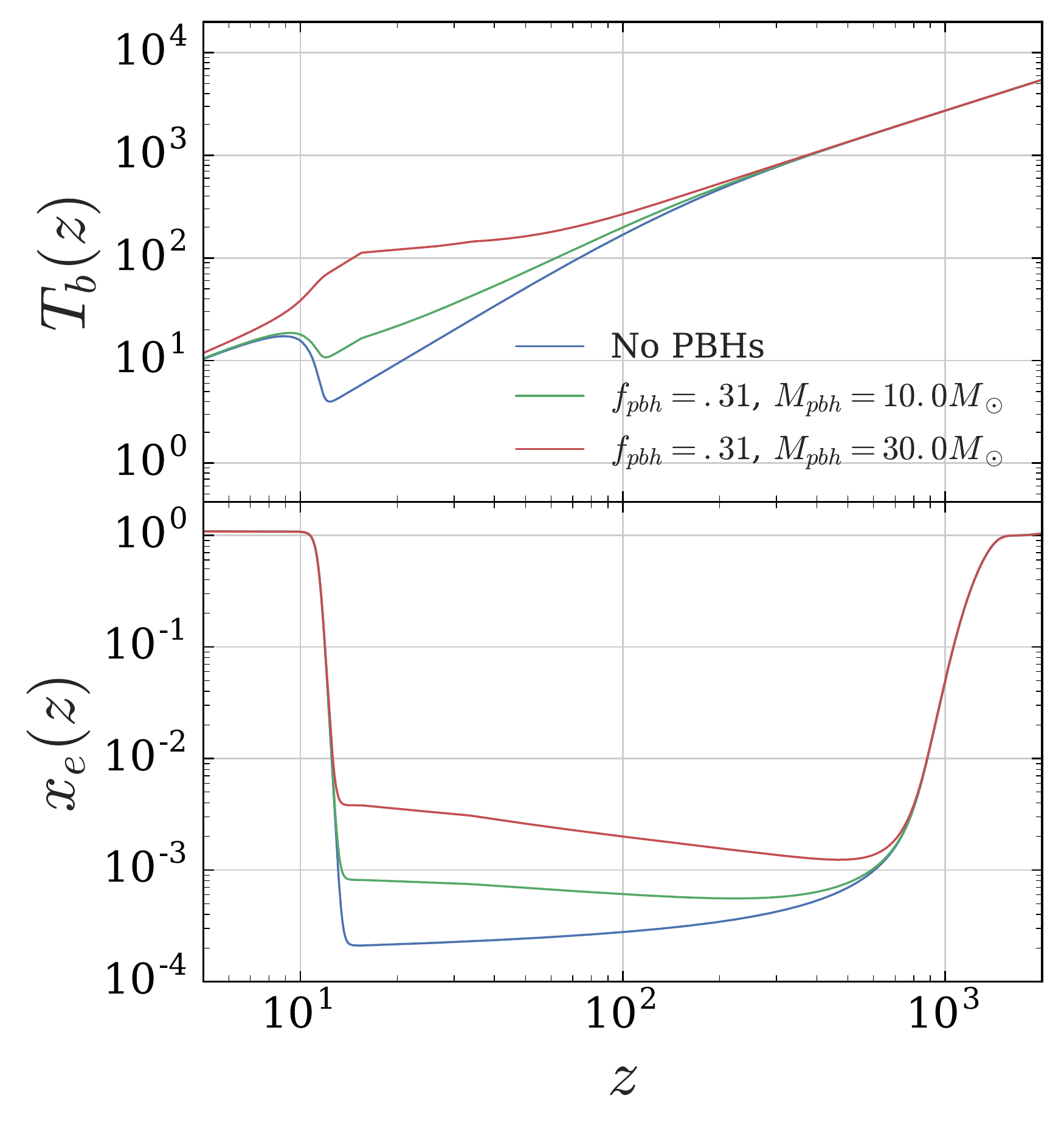}
    \caption{Effects of PBH on cosmic ionization and temperature history, as computed by CLASS.}
    \label{fig_effects}
\end{figure}

\subsection{Energy Deposition and Ionization}

In general, energy from the accreting black holes will not be deposited immediately from the source into the IGM (i.e. "on the spot") as the optical depth during the cosmic dark ages is low. We can write the energy deposited from a given emitted wavelength, $\nu_z$, at a given redshift as;
\begin{equation}
\label{fluxeq}
\frac{d^2E_{{\mbox{\scriptsize\nul}}}^{\mathrm{tot}}(z_0)}{dVd\nu}=\int_{z_{\mathrm{ol}}}^{z_0}dz 
\frac{d^3E_{{\mbox{\scriptsize\nuz}}}(z)}{dVd{\nu}dt}
e^{-\tau(z,z_0)}
\frac{dt}{dz}.
\end{equation}

%Existing constraints already disfavor high ionization fraction of the IGM at high redshift,\cite{adam2016planck} indicating that we expect the dominate contribution to $\tau(z,z_0)$ to be ionized baryonic matter forming the accretion disk around the black hole. 

The energy deposited in the IGM at a given redshift $z$ will go into both heating and ionization. The physics behind how this deposition has been a topic of recent study due to its implications in constraining dark matter annihilation and decay \cite{slatyer2013energy,poulin2015dark}. Here we briefly summarize the results of these works.

We indicate the fraction of energy deposited going into heating of the intergalactic medium as $g_{\rm h}$ and that going into ionization $g_{\rm ion}$, which can be further broken down into energy used to excite helium, $g_{\rm He}$ and hydrogen, $g_{H}$. We can write the heating rate ($dT_m/dt$) from the Primordial Black Hole accretion as
\beal
\label{eq:dT_dt}
\left.\Abl{T_{\rm M}}{t}\right|_{\rm heat}
%=\frac{f_{\rm h}(z)}{\alpha(z)}\,\left.\Abl{E}{t}\right|_{\rm \chi\bar{\chi}}
=\frac{2}{3}\,\frac{g_{\rm h}(z)}{N_{\rm H}[1+f_{\rm He}+x_{\rm e}]}\,\left.\Abl{E_{\rm d}}{t}\right|_{PBH}.
\end{align}
where $N_H$ is the total number of hydrogen nuclei, and $f_{He}$
is the number of helium nuclei relative to the number of hydrogen
nuclei. We have $g_{\rm ion}(z)=g^{\rm H}_{\rm ion}(z)+g^{\rm He}_{\rm ion}(z)$, where $g^{\rm H}_{\rm ion}(z)$ and $g^{\rm He}_{\rm ion}(z)$ are the partial contributions of hydrogen and helium, respectively.

We can relate the ionization energy deposited at a given redshift to the number density of hydrogen atoms in a given ionization state \cite{chluba2010dmann}. For hydrogen and helium, with relative fraction $f_{\rm HE}$, one has,
\bsub
\label{eq:dN_dt_i}
\beal
\label{eq:dN_dt_i_a}
\left.\Abl{N^{\ion{H}{i}}_{\rm 1s}}{t}\right|_{\rm i}
&=-\frac{1}{1+f_{\rm He}}\,\frac{g^{\rm H}_{\rm ion}(z)}{E^{\ion{H}{i}}_{\rm ion}}\,\left.\Abl{E_{\rm d}}{t}\right|_{\rm PBH}
\\
\label{eq:dN_dt_i_b}
\left.\Abl{N^{\ion{He}{i}}_{\rm 1s}}{t}\right|_{\rm i}
&=-\frac{f_{\rm He}}{1+f_{\rm He}}\,\frac{g^{\rm He}_{\rm ion}(z)}{E^{\ion{He}{i}}_{\rm ion}}\,\left.\Abl{E_{\rm d}}{t}\right|_{\rm PBH},
\end{align}
\esub
where $E^{\ion{H}{i}}_{\rm ion}=13.6\,$eV and $E^{\ion{He}{i}}_{\rm ion}=24.6\,$eV are the ionization potentials of hydrogen and helium, respectively. The coefficients $g$ have been found by \cite{slatyer2013energy} and implemented in our Boltzmann solver code, CLASS.

This simple approach provides a strict lower-bound to the ionization fraction created by primordial black hole accretion. It is possible that a combination of lower energy photons can lead to additional ionization fraction through successive excitations.

Other works that have studied the effect of primordial black holes on ionization history \cite{ricotti2008effect,chen2016constraint} find much stronger effects at re-ionization. These works model the structure of formation and find that the accretion rate would increase dramatically at late times resulting in early reionization. As shown in Figure \ref{fig_effects}, our results follow a different behavior and we find no particular change in the re-ionization history except at very high black hole mass.  As discussed in Section \ref{other_effects}, the exact behavior of the dark matter halos is model dependent but would act to increase energy output and thereby increase ionization fraction. We can therefore use our model as a strict lower-bound on the ionization caused by PBHs.

%Structure of formation could change this description, however the fraction of dark matter in halos ($M_{halo}>10^9M_\odot$) is relatively small, <1\% at z=10, so most primordial black holes are not in particularly strong over densities and shouldn't have significant changes in their accretion dynamics.

%-----------
\subsection{Effects of Ionization History on CMB Powerspectra}
\label{effects_on_cmb}

%\begin{figure}
%    \includegraphics[width=0.5\textwidth]{./plots/pbh_chisquared.pdf}
%    \caption{Comparison of the Planck Best Fit cosmology with the best fit along our borderline ($f_{pbh}=0.03$, $M_{pbh}=10$). The majority of constraining power is at low $l$, where the polarization signal is significantly effected by the ionization fraction at high $z$.}
%    \label{fig_chi}
%\end{figure}

The effect of ionization history was originally discussed in \cite{dodelson1993re} and further developed in \cite{zaldarriaga1997polarization}. There are two main effects; (a) damping of anisotropy and (b) low-$l$ polarization effects.

(a) Ionized regions have free-electrons which Thomson-scatter CMB photons. We expect a damping of the temperature proportional to $e^{-\tau}$ for all scales smaller than the horizon size. In general this effect is highly degenerate with the amplitude of the primordial power spectra, $A_s$. 

(b) Change in low $l$ polarization;  scattered radiation from a free electrons in a quadrupole radiation field will be linearly polarized. This will induce an increase in power at the $l$ associated with the horizon size at that given red-shift. In the context of reionization, this signal is sometimes referred to as the reionization ``bump.” As this signal will be at $l<30$, it is noisy due to foregrounds and inherit limits from cosmic variance.

\section{Constraints from CMB Analysis}
\label{CMB_An}

\subsection{Overview of Model}

To briefly summarize assumptions going into our model:

\begin{enumerate}[label=(\alph*)]
\item all accretion is spherically symmetric (i.e. Bondi accretion);
%\item assume a baseline duty rate of $f_{duty}=0.02$;
\item do not include any electromagnetic perturbative effects in the accretion zone;
\item no growth of structure or formation of halos around the primordial black holes;
\item no mergers of black holes or growth of the black holes via accretion;
\item no on-the-spot approximation, allowing energy to emitted at a given redshift to be absorbed at a lower redshift.
\end{enumerate}

Weakening any of these assumptions would lead to increased energy output of the PBHs, increased ionization fraction, and larger effects on the Cosmic Microwave Background.

\subsection{Analysis}

We implement our Boltzmann code into CLASS \cite{blas2011cosmic,lesgourgues2011cosmic}, using Cosmo++ \cite{aslanyan2014cosmo++} with the MultiNest algorithm \cite{feroz2013importance} to perform a maximum likelihood analysis over our parameter space. We vary the standard $\Lambda$CDM model, $( \Omega_{b},\Omega_{cdm}, h, \tau, A_s, n_s)$, as well as a nuisance parameter of Planck's Amplitude, $A_{planck}$, (with prior $1.000 \pm 0.0025$), in addition to our PBH distribution parameters, either $(f_{pbh}, M_{pbh})$ or $(M_{c}, \sigma)$.

For our data we use Planck 2015 {\tt Plik} likihood code \cite{aghanim2015planck} in the TT spectrum up to $l_{max} = 2508$ and in the TE and EE spectra up to $l_{max} = 1996$. We include the effects of lensing using reconstruction from the SMICA temperature and polarization maps. Low $l$ polarization data ($l<30$) comes from WMAP9 analysis as the Planck HFI results \cite{adam2016planck} were not publicly available as of the preparation of this paper. This dataset and associated {\tt lollipop} (the LOw-$\ell$ LIkelihood on POlarized Power-spectra) analysis code would be useful in helping constrain ionization history of the universe as discussed in Section \ref{effects_on_cmb}.

We also include large scale structure data from bayronic acoustic oscillations, by including 6dFGS, SDSS-MGS and BOSS- LOWZ BAO measurements of DV/rdrag and the CMASS-DR11 anisotropic BAO measurements via the implementation in Cosmo++ \cite{aghanim2015planck}. While this data-set doesn't get directly affected by changes in ionization history, it provides an independent probe of cosmological parameters and it helps break some of the degeneracies that affect the CMB powerspectrum. \cite{eisenstein2007improving}

One of the aspects of the $\Lambda$CDM model is the degeneracy between optical depth to the CMB, $\tau$, the primordial power spectrum amplitude, $A_s$, and spectral index, $n_s$. A dataset that could help break this degeneracy is cluster mass which provides a direct probe of local $\sigma_8$ which is affected by $A_s$ and $n_s$ but not by $\tau$. However, these cluster mass measurements are very sensitive to gas physics within the cluster and have high uncertainty \cite{von2014robust, mantz2015weighing}. Due to this uncertainty, we do not not use this data-set in this work but do note that this would be an interesting avenue to pursue in future work.

\begin{figure}
    \includegraphics[width=0.50\textwidth]{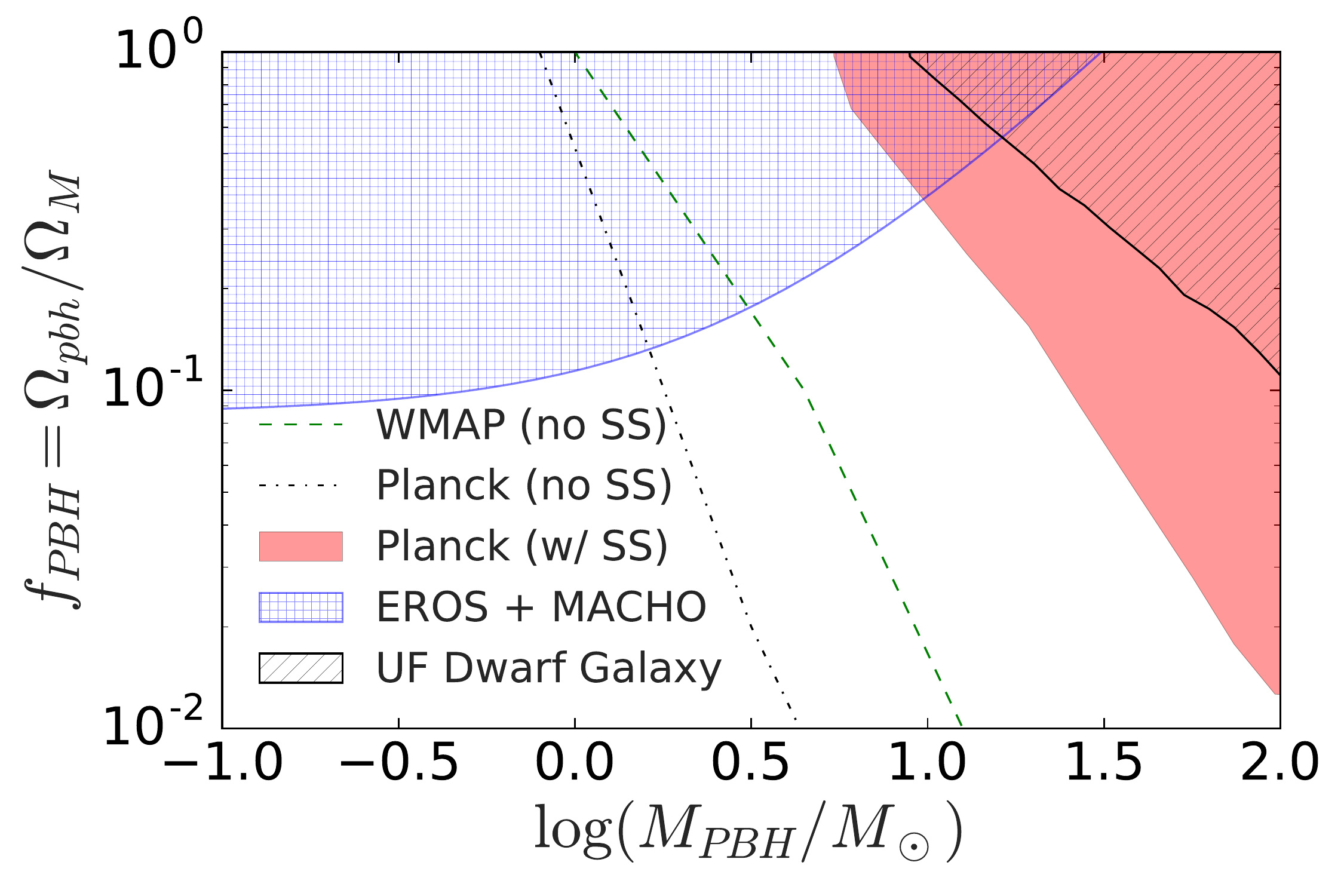}
    \caption{Comparison of constraints of this work (Planck) with those from \cite{ricotti2008effect} (WMAP), the EROS+MACHO microlensing survey \cite{alcock1998eros}, and the heating of Ultra Faint Dwarf Galaxies \cite{brandt2016constraints}. We show constraints including and not including the effect of supersonic streaming (SS). For the Dwarf Galaxy constraint, we show the constraint from the Eridanus II cluster assume an age of 3 Gyr.}
    \label{fig_con_baseline}
\end{figure}

\begin{figure}
   \includegraphics[width=0.50\textwidth]{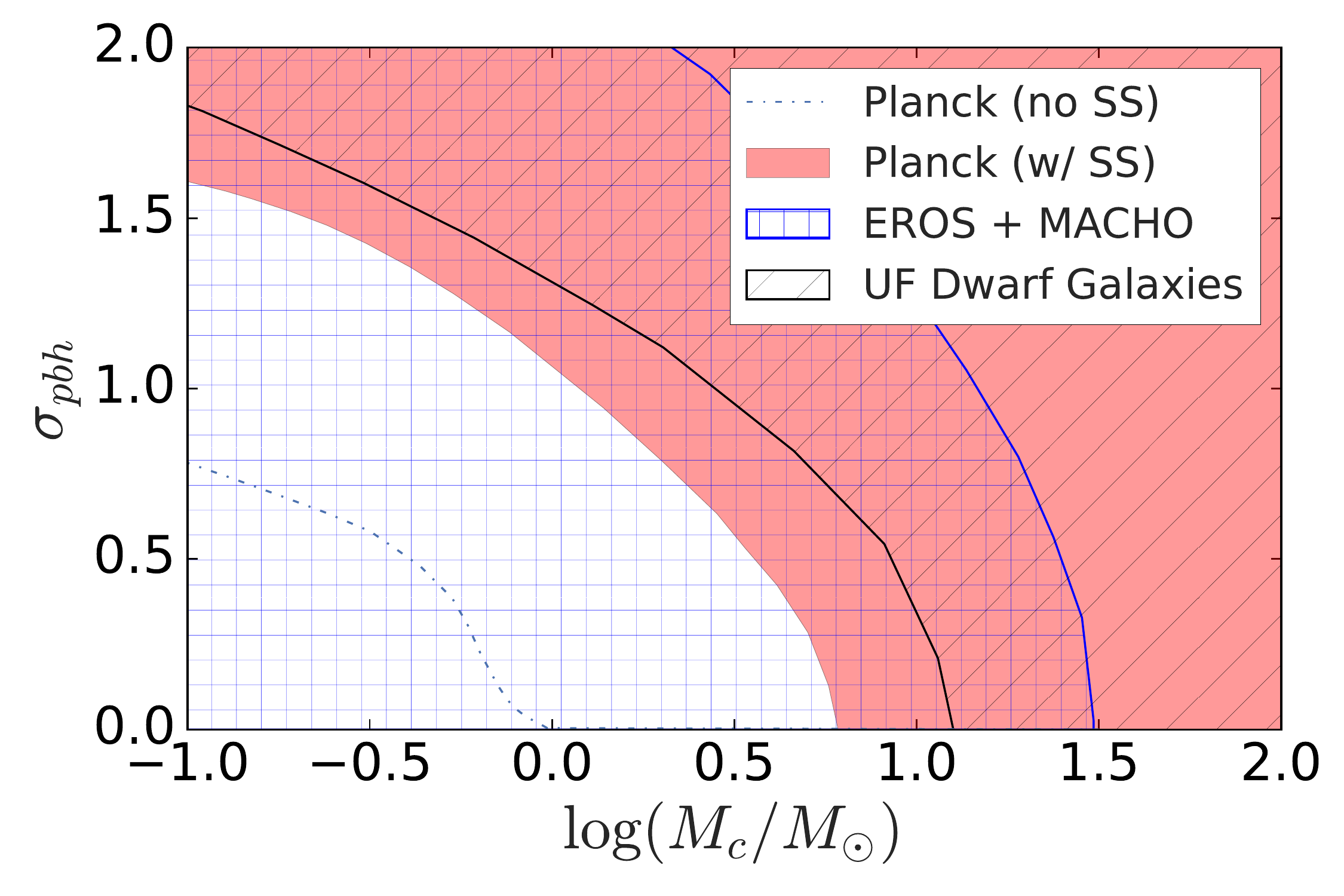}
    \caption{Comparison of constraints of this work (Planck) with those from \cite{green2016microlensing} using data from the EROS+MACHO microlensing survey \cite{alcock1998eros}, and the heating of Ultra Faint Dwarf Galaxies \cite{brandt2016constraints}.}
    \label{fig_con_extended}
\end{figure}

\section{Discussion}
\label{Discussion}

In Figure \ref{fig_con_baseline}, we show the two sigma constraints on the primordial black holes under the baseline model and in Figure \ref{fig_con_extended} we show the two sigma constraints on an extended mass function. We find that the LIGO-detection mass range ($\log{(M/M_\odot)}\approx 1.5$) is ruled out as the primary component of dark matter both with and without supersonic streaming. However, it is important to note that supersonic streaming works to significantly reduce the constraints found in \cite{ricotti2008effect, chen2016constraint} due to the strong dependence of the PBH luminosity on the relative velocity.

This constraint is consistent with those found by microlensing \cite{tisserand2007limits, alcock1998eros} and dynamic heating constraints \cite{brandt2016constraints}. While microlensing constraints have been able to rule out $30 M_\odot$ black holes with upper bound $f_{pbh}\approx 0.5$ and dynamic heating finding an upper bound of $f_{pbh}\approx 0.2$, the CMB constraints extends this constraint, finding upper limits of $f_{pbh} = .09$. 

Including structure of formation should mildly increase constraints on primordial black holes due to strong signals at low-$l$ in polarization coming from a long-duration reionization event. The constraining power of current CMB measurements on this area is somewhat limited, with some models \cite{heinrich2016complete} favoring a long re-ionization period at significant statistical significance. Recent measurements using Planck High Frequency Instrument might help shed additional constraints on this area \cite{adam2016planck}, with findings suggesting a sharp reionization transition ($\Delta z <6.8$ at 95\%). It seems likely that a new analysis with this dataset would strongly disfavor any PBH model which effects re-ionization history.

It is also possible that other cosmological probes might be able to detect signals of ionization at high redshift and constrain ionization caused by primordial black hole accretion. Of particular interest is 21cm measurements which show promise in constraining the ionization fraction out to $z=20$ \cite{mcquinn2006cosmological,liu2016constraining}. It is also likely that upcoming CMB spectroscopy, such as the PIXIE \cite{kogut2011primordial} or PRISM \cite{andre2014prism} experiments, would further restrict the parameter space of PBHs via their spectral distortion \cite{ricotti2008effect}.

A natural question to ask is whether it is possible that an allowed population of primordial black holes from recombination could have evolved through mergers or accretion into the LIGO mass-range. Even in more accretion favorable models discussed in \cite{ricotti2008effect}, $\dot{m}$ is only of order $\sim 0.0001$ at maximum for a solar mass black hole, corresponding to a total mass accretion of $(\Delta t) \dot{m} \dot{M}_{Eddington} = 10^{-4} M_{\odot}$. In the model considered in this paper, the net mass change due to accretion from recombination till today is significantly smaller, $10^{-8} M_{\odot}$.

Similarly, order of magnitude calculations done by \cite{bird2016did,sasaki2016primordial} suggest extremely small merger rates of PBHs of <1000 Gpc$^{-3}$ yr$^{-1}$. This is vanishingly small compared to the number density of dark matter PBHs of $\sim 10^{19} M_\odot/M_{pbh}$ Gpc$^{-3}$. Even with the large population of early-forming binary PBH pairs predicted by \cite{nakamura1997gravitational}, it will require multiple merger events per black hole to get a significant population of $30 M_\odot$ PBHs.

\section*{Acknowledgments}
We are extremely grateful to Grigor Aslanyan for his assistance in running Cosmo++ and discussion useful to the analysis. We also thank Surjeet Rajendran, Chirag Modi, and  Katelin Schutz for discussion useful to this analysis, with particular thanks to Katelin Shutz for pointing out the early supersonic streaming effect. BH is supported by the National Science Foundation, award number DGE 1106400. 

This research used resources of the National Energy Research Scientific Computing Center, a DOE Office of Science User Facility supported by the Office of Science of the U.S. Department of Energy under Contract No. DE-AC02-05CH11231.

\emph{Note:} During the final stages of the preparation of this draft, the analysis of \cite{ak} was released which performed a similar analysis using a new accretion feedback model. Their ``strong feedback'' results are in line with the constraints found in this paper. Adjusting our $f_{duty}$ parameter (discussed in Section \ref{Accretion_Physics}) from 1 to 0.01 roughly duplicates their ``no feedback'' case in the mass range in consideration in this work.

\bibliographystyle{apsrev4-1}
\bibliography{sample}

\end{document}